\documentclass[12pt]{article}
\usepackage{sc3conf}
\usepackage{amsfonts}
\usepackage{epic}

\begin{document}
\raggedbottom

\title{The motion of a shock wave through a non-uniform one-dimensional 
medium in the case of arbitrary equation of state}

\authors{S.A.~Serov \adref{1,2}}

\addresses{\1ad Institute of Theoretical and Mathematical Physics, 
                RFNC-VNIIEF, Sarov,
  \nextaddress \2ad e-mail: {hobo@albatross.md08.vniief.ru}.}

\maketitle

\begin{abstract}
  The derivation of the equation of one-dimensional movement 
  of a solitary shock wave is given. 
  This derivation shows, that the differential equation of movement of a 
  solitary plane shock wave in the channel with variable area, 
  is exact, if simplifying assumptions, made during 
  derivation, are realized. 
  But these assumptions in plane geometry it is possible to realize only 
  approximately; 
  situation  with spherical and cylindrical shock waves is opposite.
\end{abstract}

\section{Introduction} \label{sec:introduction}
In 1957 Chisnell \cite{chisnell57}, being based on ideas of Chester 
\cite{chester54}, has deduced the equation of one-dimensional movement of a 
solitary shock wave on substance with polytropic equation of state. 
A little bit later the Whitham \cite{whitham58} has offered "... simple rule",
which together with shock relations 
"... determines the motion of the shock wave". 
Actually Whitham was the first, who has considered the equation of 
one-dimensional movement of a solitary shock wave on substance with the 
arbitrary equation of state (see below the equation (\ref{eq14})), 
but as the approximate equation, from which for a special case of the 
polytropic equation of a state, one can derive the Chisnell equation.

\section{The derivation of the equation} \label{sec:derivation}
We shall consider one-dimensional movements of solitary shock waves. 
Movement of a shock wave is identified with movement of its front. 
One-dimensional movements of shock waves are understood as movements 
of spherically symmetric, spherically symmetric and plane shock waves. 
Shock waves are assumed solitary, i.e. it is not considered overtaking one 
shock wave by other shock wave etc. 
Naturally, the solitary shock wave is an abstraction.
Below, during the derivation of the equation of one-dimensional 
movement of a solitary plane shock wave the differential calculus will be used, 
how it was intuitively used in Newton time. 
After development of the nonstandard analysis by Robinson 
(see \cite{devis80}) such operating with infinitesimal differentials 
may be regarded as quite correct.

Following Chisnell, we shall consider movement of a \textit{plane} shock 
wave in the channel, having in some place infinitesimal jump of the 
sectional area $dA$ -- see figure (cp. it with Fig.~1 in Chisnell's article 
\cite{chisnell57}, this figure can be also compared with Fig.~1, p.~193 in 
\cite{ovsyannikov81}).
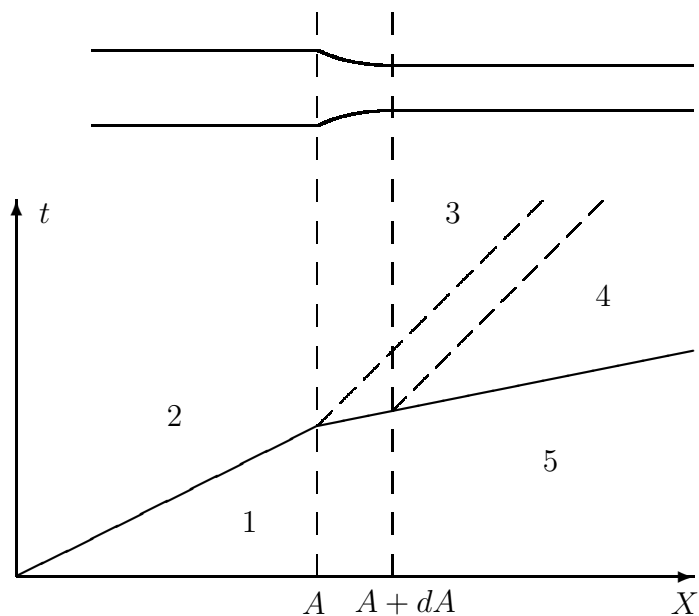
\begin{figure}[htb]
  \begin{center}
  %Created by PicEdt 1.x
%Tue Oct 15 11:56:55 MSD 2002
\unitlength 1mm
\begin{picture}(115.0,85.00)(0,0)

\linethickness{0.5mm}
\thicklines
%Line 0 1 (20.00,10.00)(20.00,60.00)
\drawline(20.00,10.00)(20.00,60.00)
\put(20.00,60.00){\vector(0,1){0.12}}
%End Line

%Line 0 1 (20.00,10.00)(110.00,10.00)
\drawline(20.00,10.00)(110.00,10.00)
\put(110.00,10.00){\vector(1,-0){0.12}}
%End Line

%Line 0 0 (30.00,70.00)(60.00,70.00)
\drawline(30.00,70.00)(60.00,70.00)
%End Line

%Line 0 0 (70.00,72.00)(110.00,72.00)
\drawline(70.00,72.00)(110.00,72.00)
%End Line

%Line 0 0 (30.00,80.00)(60.00,80.00)
\drawline(30.00,80.00)(60.00,80.00)
%End Line

%Line 0 0 (70.00,78.00)(110.00,78.00)
\drawline(70.00,78.00)(110.00,78.00)
%End Line

%Line 0 0 (20.00,10.00)(60.00,30.00)
\drawline(20.00,10.00)(60.00,30.00)
%End Line

%Line 0 0 (60.00,30.00)(110.00,40.00)
\drawline(60.00,30.00)(110.00,40.00)
%End Line

\linethickness{0.4mm}
\thicklines
%Bezier 0 0 (60.00,70.00)(64.00,72.00)(70.00,72.00)
\qbezier[1000](60.00,70.00)(64.00,72.00)(70.00,72.00)
%End Bezier

%Bezier 0 0 (60.00,80.00)(64.00,78.00)(70.00,78.00)
\qbezier[1000](60.00,80.00)(64.00,78.00)(70.00,78.00)
%End Bezier

\linethickness{0.15mm}
\thinlines
%Line 0 0 (60.00,10.00)(60.00,85.00)
\dashline[+15]{3}(60.00,10.00)(60.00,85.00)
%End Line

%Line 0 0 (70.00,10.00)(70.00,85.00)
\dashline[+15]{3}(70.00,10.00)(70.00,85.00)
%End Line

%Line 0 0 (60.00,30.00)(110.00,50.00)
\dashline[+30]{3}(60.00,30.00)(90.00,60.00)
%End Line

%Line 0 0 (70.00,32.00)(110.00,48.00)
\dashline[+30]{3}(70.00,32.00)(98.00,60.00)
%End Line

\put(50.00,16.00){\makebox(0.00,0.00)[bl]{1}}
\put(40.00,30.00){\makebox(0.00,0.00)[bl]{2}}
\put(77.00,57.00){\makebox(0.00,0.00)[bl]{3}}
\put(97.00,46.00){\makebox(0.00,0.00)[bl]{4}}
\put(90.00,24.00){\makebox(0.00,0.00)[bl]{5}}

\put(23.00,57.00){\makebox(0.00,0.00)[bl]{$t$}}
\put(107.00,5.00){\makebox(0.00,0.00)[bl]{$X$}}
\put(58.00,5.00){\makebox(0.00,0.00)[bl]{$A$}}
\put(65.00,5.00){\makebox(0.00,0.00)[bl]{$A+dA$}}

\end{picture}
    \caption{The shock wave, separating region 1, 2, is incident on small 
             change in the area of a channel from $A$  to $A+dA$. The resulting 
             transmitted shock separates regions 4, 5. Regions 3, 4 are 
             separated by simple Riemann wave. 
             The shape of the channel is shown on the top of the figure.
            }\label{fig:figure1}
  \end{center}
\end{figure}
Let's choose the system of coordinates connected to the channel, which axis 
$X$ we shall direct along the channel. We shall assume, that walls of 
the channel are indefinitely rigid, so it is possible to neglect interaction 
of a shock wave with them, the plane front of a shock wave is perpendicular 
to walls of the channel etc. (i.e. it is supposed, that it is possible to 
neglect the three-dimensional effects arising at movement of a shock wave), 
the substance before front of a shock wave is homogeneous and is motionless 
relatively the channel (at presence of the area change last assumption is 
implicitly containing in the assumption of homogeneity of substance before 
front of a shock wave). As it is usually accepted, by an index "$_{0} $" 
we shall mark parameters of substance before front of a shock wave. As 
one-dimensional movement is considered, all vectors can be identified with 
their projections to axis $X$. Thus: $u_{0} = 0$, $\rho _{0} $, 
$p_{0} $, $\varepsilon _{0} $, $s_{0} $ -- respectively, speed ($\equiv $ 
projection of speed on axis $X$), density, pressure, specific 
(referred to a mass unit) internal energy, specific (referred to a mass 
unit) entropy of substance before front of a shock wave. 
As the parameter, defining strength of a shock wave, we shall take:
\begin{equation}
\label{eq1}
\sigma = \frac{{\rho} }{{\rho _{0}} }.
\end{equation}
-- the ratio of density of substance behind front of a shock 
wave to density of substance before front of a shock wave. It is 
universal, dimensionless, finite parameter. Chisnell as the parameter, 
describing strength of a shock wave, have used the ratio of pressure: 
$z = {p \mathord{\left/
 {\vphantom {p {p_0 }}} \right.
 \kern-\nulldelimiterspace} {p_0 }}
$, -- that can not be used, for example, for some equations of a state of 
substance, when $p_0 = 0$. 
In one-dimensional case assignment of $\sigma $ (or $z$), 
with known parameters of substance before the front of the shock wave 
$u_{0} = 0$, $\rho _{0} $, $p_{0} $
completely determines parameters of substance behind front of a 
shock wave and speed of movement of shock wave $D$ -- see 
theorem 5.5 in \cite{ovsyannikov81}.

On Fig.~1 the situation is schematically represented, when the 
shock wave, moving from the left to the right in positive direction of axis 
$X$ ($D > 0$), will pass through jump of area of the channel $dA$. 
Arising gas-dynamic break splits into simple Riemann $r$-wave 
(according to definition in \S~16, \cite{ovsyannikov81})
moving back from the front of the shock wave and shock wave with changed 
on $d \sigma $ strength, moving to the right (see \S~17 in 
\cite{ovsyannikov81}). 
For initial shock wave moving from right to left 
($D < 0$) instead of $r$-wave we should have Riemann $l$-wave
 -- theorem 16.2 in \cite{ovsyannikov81}.  
Situation, represented on Fig.~1 corresponds to supersonic motion 
behind front of the shock wave $\left| u \right| > c$, where $c$ -- 
speed of a sound behind front of a shock wave. Following reasonings 
are correct without change and in a subsonic case -- $\left| u \right| < c$.

Changes of speed $u$ and pressure $p$ behind front of the shock wave, 
corresponding to change of shock wave strength after passage of 
the area change, are "compensated" by changes of 
speed $d_{A} u$ and pressure $d_{A} p$ at adiabatic flowing of substance 
behind front of shock wave through the jump of the sectional area and 
by changes of speed and pressure in Riemann wave (for Riemann $r$-wave, 
when $D > 0$, index "$_{+} $" will be used, for 
$l$-wave -- index "$_{-} $"). Thus, we have the system 
of the equations:
\begin{equation}
\label{eq2}
\frac{{du}}{{d\sigma} } \cdot d\sigma = d_{A} u + d_{ \pm}  u,
\end{equation}
\begin{equation}
\label{eq3}
\frac{{dp}}{{d\sigma} } \cdot d\sigma = d_{A} p + d_{ \pm}  p.
\end{equation}
At adiabatic flowing of substance behind front of the shock wave through the 
jump of the sectional area $dA$ following relations must be satisfied:
\begin{equation}
\label{eq4}
d_{A} \left( {A\rho u} \right) = 0,
\end{equation}
\begin{equation}
\label{eq5}
d_{A} \left( {\frac{{u^{2}}}{{2}} + \varepsilon + \frac{{p}}{{\rho} }} 
\right) = 0,
\end{equation}
\begin{equation}
\label{eq6}
d_{A} s = 0.
\end{equation}
First of them expresses law of conservation of mass, the second is Bernoulli 
equation, the last -- adiabatic condition. The adiabatic condition implies 
equalities:
\begin{equation}
\label{eq7}
d_{A} p = c^{2} \cdot d_{A} \rho ,
\end{equation}
\begin{equation}
\label{eq8}
d_{A} \varepsilon = \frac{{p}}{{\rho ^{2}}} \cdot d_{A} \rho .
\end{equation}
Using (\ref{eq7}) and (\ref{eq8}), from (\ref{eq4})-(\ref{eq6}) we receive system of the equations:
\begin{equation}
\label{eq9}
d_{A} u = - \frac{{c^{2}}}{{u}} \cdot \frac{{d_{A} \rho} }{{\rho} },
\end{equation}
\begin{equation}
\label{eq10}
d_{A} p = c^{2} \cdot d_{A} \rho ,
\end{equation}
\begin{equation}
\label{eq11}
 - \frac{{dA}}{{A}} = \frac{{d_{A} \rho} }{{\rho} } \cdot \left( {1 - 
\frac{{c^{2}}}{{u^{2}}}} \right).
\end{equation}
The system of the equations (\ref{eq9})-(\ref{eq11}) completely determines 
changes of parameters of substance at its motion through the jump of the 
sectional area of the channel. 

In Riemann wave the changes of parameters of substance are in accord with 
relations:
\begin{equation}
\label{eq12}
d_{ \pm}  u \pm \frac{{d_{ \pm}  p}}{{\rho c}} = 0,
\end{equation}
\begin{equation}
\label{eq13}
d_{ \pm}  p = c^{2} \cdot d_{ \pm}  \rho .
\end{equation}
The equation (\ref{eq12}) follows from definition of Riemann invariants and the 
equation (\ref{eq13}) follows from adiabatic condition for movement in Riemann wave.

Using (\ref{eq9})-(\ref{eq11}) and (\ref{eq12})-(\ref{eq13}), 
from (\ref{eq2})-(\ref{eq3}) we receive the equation:
\begin{equation}
\label{eq14}
 - \frac{{dA}}{{A}} = \left( {\frac{{1}}{{u}} \pm \frac{{1}}{{c}}} \right) 
\cdot \left( {\frac{{du}}{{d\sigma} } \pm \frac{{1}}{{\rho c}} \cdot 
\frac{{dp}}{{d\sigma} }} \right) \cdot d\sigma .
\end{equation}
For a shock wave, moving from left to right ($D > 0$), in the right part 
(\ref{eq14}) it is necessary to take signs "$ + $". If $D < 0$ (the shock wave 
moves from right to left), in the right part of the equation (\ref{eq14}) it is 
necessary to take signs "$ - $". It is interesting to note, that the right 
part of the equation (\ref{eq14}) is a product of a combination of speed of 
substance and speed of a sound on differential of appropriate Riemann 
invariant.

For the channel with arbitrary (smooth) dependence of the area of section on 
coordinate $x$ -- $A\left( {x} \right)$, in the assumption of plane 
symmetry of considered movement, it is possible to pass from differential 
(\ref{eq14}) to integrated equality:
\begin{equation}
\label{eq15}
 - \int\limits_{A_{i}} ^{A} {\frac{{d{A}'}}{{{A}'}} = \int\limits_{\sigma 
_{i}} ^{\sigma}  {\left( {\frac{{1}}{{u}} \pm \frac{{1}}{{c}}} \right) \cdot 
\left( {\frac{{du}}{{d{\sigma} '}} \pm \frac{{1}}{{\rho c}} \cdot 
\frac{{dp}}{{d{\sigma} '}}} \right) \cdot d{\sigma} '}}  ,
\end{equation}
-- where $A_{i} $ and $\sigma _{i} $ -- values $A$ and $\sigma $ 
at some moment of time $t_{i} $.

The equation (\ref{eq15}) can be named (integrated) equation of movement of a 
solitary plane shock wave on homogeneous substance in the channel of 
variable area, as at known dependence $A\left( {x} \right)$ it determines 
speed of shock wave $D$ as implicit function of coordinate $x$ and with 
trivial equation:
\begin{equation}
\label{eq16}
\frac{{dx}}{{dt}} = D\left( {x} \right),
\end{equation}
-- allows to describe movement of a shock wave completely. The equation (\ref{eq14}) 
now, accordingly, can be named the differential equation of movement of a 
plane solitary shock wave on homogeneous substance in the channel of 
variable section.

The equation (\ref{eq14}) can be trivial generalized, if substance before front of a 
shock wave is not homogeneous. For this purpose, in equations(\ref{eq2}), (\ref{eq3}) it is 
necessary to replace differentials $\frac{{du}}{{d\sigma} } \cdot d\sigma $ 
and $\frac{{dp}}{{d\sigma} } \cdot d\sigma $ with the more general 
expressions $du$ and $dp$, because using theorem 5.5 in \cite{ovsyannikov81} 
one can write:
\begin{equation}
\label{eq17}
du = \frac{{\partial u}}{{\partial \sigma} } \cdot d\sigma + \frac{{\partial 
u}}{{\partial u_{0}} } \cdot du_{0} + \frac{{\partial u}}{{\partial \rho 
_{0}} } \cdot d\rho _{0} + \frac{{\partial u}}{{\partial p_{0}} } \cdot 
dp_{0} ,
\end{equation}
\begin{equation}
\label{eq18}
dp = \frac{{\partial p}}{{\partial \sigma} } \cdot d\sigma + \frac{{\partial 
p}}{{\partial u_{0}} } \cdot du_{0} + \frac{{\partial p}}{{\partial \rho 
_{0}} } \cdot d\rho _{0} + \frac{{\partial p}}{{\partial p_{0}} } \cdot 
dp_{0} ,
\end{equation}
-- instead of $\sigma $, $u_{0} $, $\rho _{0} $, $p_{0} $, other choice of 
parameters is possible also. After that, from (\ref{eq4})-(\ref{eq13}), we can finally 
receive following differential equation of movement of a solitary shock 
wave:
\begin{equation}
\label{eq19}
 - \frac{{dA}}{{A}} = \left( {\frac{{1}}{{u}} \pm \frac{{1}}{{c}}} \right) 
\cdot \left( {du \pm \frac{{1}}{{\rho c}} \cdot dp} \right).
\end{equation}
It is supposed, that appearing in (\ref{eq17})-(\ref{eq19}) functions $A\left( {x} 
\right)$, $u_{0} \left( {x} \right)$, $\rho _{0} \left( {x} 
\right)$, $p_{0} \left( {x} \right)$ are smooth. For the first time 
\textit{similar} generalization was proposed by Vakhrameev -- see 
\cite{vakhrameev66}, \cite{vakhrameev90}.

\section{Chisnell's solution} \label{sec:solution}
Chisnell in article \cite{chisnell57} has derived analogue of the equation (\ref{eq14}) for a 
special case of polytropic equation of state:
\begin{equation}
\label{eq20}
\begin{array}{l}
 - \frac{{1}}{{A}} \cdot \frac{{dA\left( {z} \right)}}{{dz}} = 
\frac{{1}}{{\gamma z}} + \frac{{1}}{{\left( {z - 1} \right)}} - 
\frac{{\left( {\gamma + 1} \right)}}{{2\left\{ {\left( {\gamma + 1} \right)z 
+ \left( {\gamma - 1} \right)} \right\}}} + \\ 
 + \left[ {\frac{{2}}{{\gamma z\left\{ {\left( {\gamma - 1} \right)z + 
\left( {\gamma + 1} \right)} \right\}}}} \right]^{\frac{{1}}{{2}}} \times \\ 
 \times \left[ {1 - \frac{{\left( {\gamma + 1} \right) \cdot \left( {z - 1} 
\right)}}{{2\left\{ {\left( {\gamma + 1} \right)z + \left( {\gamma - 1} 
\right)} \right\}}} + \frac{{\left( {\gamma - 1} \right)z + \left( {\gamma + 
1} \right)}}{{2\left( {z - 1} \right)}}} \right]. \\ 
 \end{array}
\end{equation}
Here $\gamma $ is adiabatic exponent for polytropic gas, $z = {{p} 
\mathord{\left/ {\vphantom {{p} {p_{0}} }} \right. 
\kern-\nulldelimiterspace} {p_{0}} }$. The parameter $z$, describing 
force of a shock wave, is determined correctly, as for the polytropic 
equations of a state $\rho _{0} \ne 0$ implies $p_{0} \ne 0$, if only the 
temperature is not equal to absolute zero.

The indefinite integral of the equation (\ref{eq20}) found by Chisnell, looks like:
\begin{equation}
\label{eq21}
Af\left( {z} \right) = constant.
\end{equation}
Where
\begin{equation}
\label{eq22}
\begin{array}{l}
 f\left( {z} \right) = z^{\frac{{1}}{{\gamma} }}\left( {z - 1} \right) \cdot 
\left( {z + \frac{{\gamma - 1}}{{\gamma + 1}}} \right)^{ - \frac{{1}}{{2}}} 
\times \\ 
 \times \left[ {\frac{{1 + \left\{ {1 + \frac{{\left( {\gamma + 1} 
\right)}}{{\left( {\gamma - 1} \right)z}}} \right\}^{ - 
\frac{{1}}{{2}}}}}{{1 - \left\{ {1 + \frac{{\left( {\gamma + 1} 
\right)}}{{\left( {\gamma - 1} \right)z}}} \right\}^{ - \frac{{1}}{{2}}}}}} 
\right]^{\sqrt {\frac{{\gamma} }{{2\left( {\gamma - 1} \right)}}}}  \times 
\\ 
 \times \left[ {\frac{{\left\{ {1 + \frac{{\left( {\gamma + 1} 
\right)}}{{\left( {\gamma - 1} \right)z}}} \right\}^{ - \frac{{1}}{{2}}} - 
\left( {\frac{{\gamma - 1}}{{2\gamma} }} 
\right)^{\frac{{1}}{{2}}}}}{{\left\{ {1 + \frac{{\left( {\gamma + 1} 
\right)}}{{\left( {\gamma - 1} \right)z}}} \right\}^{ - \frac{{1}}{{2}}} + 
\left( {\frac{{\gamma - 1}}{{2\gamma} }} \right)^{\frac{{1}}{{2}}}}}} 
\right] \times \\ 
 \times exp\left[ {\left( {\frac{{2}}{{\gamma - 1}}} 
\right)^{\frac{{1}}{{2}}} \cdot arctan\left\{ {\frac{{2}}{{\left( {\gamma - 
1} \right)}} \cdot \left( {\frac{{\gamma z}}{{z + \frac{{\gamma + 
1}}{{\gamma - 1}}}}} \right)} \right\}} \right]. \\ 
 \end{array}
\end{equation}
Let's look, how in case of polytropic equation of a state the equation 
(\ref{eq20}) can be derived from the equation (\ref{eq14}). For definiteness 
we shall consider shock wave, moving in a positive direction of axis $X$ ($D > 0$). Then 
we take signs "$ + $" in (\ref{eq14}):
\begin{equation}
\label{eq23}
 - \frac{{1}}{{A}} \cdot \frac{{dA}}{{dz}} = \left( {\frac{{1}}{{u}} + 
\frac{{1}}{{c}}} \right) \cdot \left( {\frac{{du}}{{dz}} + \frac{{1}}{{\rho 
c}} \cdot \frac{{dp}}{{dz}}} \right).
\end{equation}
Parameters of substance behind front of a shock wave are determined by 
Rankine-Hugoniot shock relations, which for polytropic equations of a state 
can be written in the following form:
\begin{equation}
\label{eq24}
u\left( {z} \right) = \left( {z - 1} \right) \cdot \left[ {\frac{{2p_{0} 
}}{{\rho _{0} \left\{ {\left( {\gamma + 1} \right)z + \left( {\gamma - 1} 
\right)} \right\}}}} \right]^{\frac{{1}}{{2}}},
\end{equation}
\begin{equation}
\label{eq25}
\rho \left( {z} \right) = \rho _{0} \cdot \frac{{\left( {\gamma + 1} 
\right)z + \left( {\gamma - 1} \right)}}{{\left( {\gamma - 1} \right)z + 
\left( {\gamma + 1} \right)}},
\end{equation}
\begin{equation}
\label{eq26}
c\left( {z} \right) = \sqrt {\frac{{\gamma p}}{{\rho} }} = \sqrt 
{\frac{{\gamma zp_{0}} }{{\rho _{0}} } \cdot \frac{{\left( {\gamma - 1} 
\right)z + \left( {\gamma + 1} \right)}}{{\left( {\gamma + 1} \right)z + 
\left( {\gamma - 1} \right)}}} .
\end{equation}
Differentiation of (\ref{eq24}) gives:
\begin{equation}
\label{eq27}
\frac{{du}}{{dz}} = \left[ {\frac{{2p_{0}} }{{\rho _{0} \left\{ {\left( 
{\gamma + 1} \right)z + \left( {\gamma - 1} \right)} \right\}}}} 
\right]^{\frac{{1}}{{2}}} \cdot \left[ {1 - \frac{{\left( {z - 1} \right) 
\cdot \left( {\gamma + 1} \right)}}{{2\left\{ {\left( {\gamma + 1} \right)z 
+ \left( {\gamma - 1} \right)} \right\}}}} \right].
\end{equation}
Substituting (\ref{eq24})-(\ref{eq27}) in (\ref{eq23}), we receive 
Chisnell equation (\ref{eq20}).

\section{Comparison of different solutions} \label{sec:comparing}
One of the most important examples of one-dimensional movements of shock 
waves is movement on homogeneous substance solitary converging spherically 
symmetric shock wave. Choosing small enough solid angle with top in the 
centre of symmetry of a shock wave, we can see, that it is possible to 
consider movement of a shock wave inside such solid angle, as movement in 
the channel with the variable sectional area ($A = constant \cdot r^{\alpha 
}$, $\alpha = 2$, $r$ -- radius of front of a shock wave in spherical system 
of coordinates with the beginning in the centre of symmetry of a shock 
wave), and all assumptions of section 2 are precisely carried out. Therefore 
movement of spherically symmetric shock wave in spherical system of 
coordinates, which beginning coincides with the centre of symmetry of a 
shock wave, is described by the equation (see the equation (\ref{eq15})):
\begin{equation}
\label{eq28}
 - \alpha \cdot ln\frac{{r}}{{r_{i}} } = \int\limits_{\sigma _{i}} ^{\sigma 
} {\left( {\frac{{1}}{{u}} \pm \frac{{1}}{{c}}} \right) \cdot \left( 
{\frac{{du}}{{d{\sigma} '}} \pm \frac{{1}}{{\rho c}} \cdot 
\frac{{dp}}{{d{\sigma} '}}} \right) \cdot d{\sigma} '} .
\end{equation}
Similar reasonings show, that the equation (\ref{eq28}) describes also 
movement of solitary converging cylindrically symmetric shock waves 
($\alpha = 1$, $r$ -- 
radius of front of a shock wave in cylindrical system of coordinates with 
the beginning on an axis of symmetry of a shock wave). Signs in the right 
part of the equation (\ref{eq28}) are taken according to agreements in 
Sec.~\ref{sec:derivation}).

The finding of the analytical solution of the equation (\ref{eq28}) for the 
concrete equation of state of substance can be not trivial problem -- 
see Chisnell's solution (\ref{eq21})-(\ref{eq22}). However, with the help of 
asymptotic analysis of functions of real variable 
(see, for example, chapter 5 in \cite{bourbaki65}) it is often 
easily to receive asymptotic solution (at $r \to 0$) of the 
equation (\ref{eq28}), that is also interesting enough. 
Thus received asymptotic 
solution of the equation (\ref{eq28}), generally speaking, are 
\textit{logarithmically equivalent} to the true solution -- see definition 5, 
\S~1, chapters 5 in \cite{bourbaki65}. 
It is dictated by structure of the right 
part of the equation (\ref{eq28}). 
Analytical Chisnell solution for polytropic 
equations of state allows to receive \textit{strongly equivalent} solution 
of the equation (\ref{eq28}) 
(see definition 4, \S~1, chapters 5 in \cite{bourbaki65}):
\begin{equation}
\label{eq29}
\frac{{p}}{{p_{i}} }\sim \left( {\frac{{r}}{{r_{i}} }} \right)^{ - \alpha \nu 
}.
\end{equation}
Here $p_{i} $ -- pressure behind front of the shock wave, which is taking 
place at some moment of time $t_{i} $ on radius $r_{i} $ (it is supposed, 
that $p_{i} > p_{0} $); $\alpha $ -- is determined above, $\nu $ is 
expressed by the formula:
\begin{equation}
\label{eq30}
\nu = \left[ {\frac{{\gamma + 2}}{{2\gamma} } + \frac{{1}}{{2}} \cdot \sqrt 
{\frac{{2\gamma} }{{\gamma - 1}}}}  \right]^{ - 1}.
\end{equation}
In the table exponent $\alpha \nu $ are given for 
$\gamma = {{5} 
\mathord{\left/ {\vphantom {{5} {3}}} \right. \kern-\nulldelimiterspace} 
{3}},\  {{7} \mathord{\left/ {\vphantom {{7} {5}}} \right. 
\kern-\nulldelimiterspace} {5}},\  {{6} \mathord{\left/ {\vphantom {{6} 
{5}}} \right. \kern-\nulldelimiterspace} {5}}$ 
together with the corresponding values from self-similar solutions: 
Guderley, Butler -- see \cite{chisnell57}, \cite{guderley42}, \cite{butler54}, 
Landau and Stanukovich -- see \S~64 in \cite{stanukovich55}, 
see also \cite{brushlinskii63}.
%\setlength\LTcapwidth{3.7in}
%\begin{longtable}{|c|c|c|c|c|}
\begin{table}[htb]
\begin{center}
\caption{Values of the module of exponent 
according to calculations of different authors.}\label{tab:table1}
\begin{tabular}{|c|c|c|c|c|}
\hline
& \multicolumn{4}{|c|}{} \\[-2.8mm] 
& \multicolumn{4}{|c|}{Cylindrical wave ($\alpha = 1$)} \\[1.4mm]
\cline{2-5}
& & & & \\[-2.8mm]
& Chisnell& Butler& Guderley& Landau, \\
& & & & Stanukovich \\[1.4mm]
\hline
& & & & \\[-2.8mm]
$\gamma = {6 \over 5}$& 0,326223& 0,322441& & \\[1.4mm]
\hline
& & & & \\[-2.8mm]
$\gamma = {7 \over 5}$& 0,394141& 0,394589& 0,396& 0,398 \\[1.4mm]
\hline 
& & & & \\[-2.8mm]
$\gamma = {5 \over 3}$& 0,450850& 0,452108& & \\[1.4mm]
\hline
& \multicolumn{4}{|c|}{} \\[-2.8mm]
& \multicolumn{4}{|c|}{Spherical wave ($\alpha = 2$)} \\[1.4mm]
\cline{2-5}
& & & & \\[-2.8mm]
& Chisnell& Butler& Guderley& Landau, \\
& & & & Stanukovich \\[1.4mm]
\hline
& & & & \\[-2.8mm]
$\gamma = {6 \over 5}$& 0,652447& 0,641513& & \\[1.4mm]
\hline
& & & & \\[-2.8mm]
$\gamma = {7 \over 5}$& 0,788283& 0,788728& & 0,789 \\[1.4mm]
\hline 
& & & & \\[-2.8mm]
$\gamma = {5 \over 3}$& 0,901699& 0,905385& & \\[1.4mm]
\hline
\end{tabular}
\end{center}
\end{table}
%\end{longtable}

It is visible, that the accordance is very good. 
Small distinction of values 
could be tried to explain, how it tried to make Chisnell \cite{chisnell57}, 
that at derivation of the equation (\ref{eq14}) possible change 
$d^{2}\sigma $ of strength of shock wave, caused by reflection of the simple 
Riemann wave, which has arisen at jump of the sectional area of the channel 
$dA_{2} $, from earlier 
arisen distortion (at jump of the sectional area of the channel $dA_{1} $) 
behind front of a shock wave. But $d^{2}\sigma $ must have second 
infinitesimal order, because $d^{2}\sigma $ is bilinear function of $dA_{1} 
$ and $dA_{2} $ -- $d^{2}\sigma \approx dA_{1} \cdot dA_{2} $ ($d^{2}\sigma $ 
can have first infinitesimal order for divergent shock waves).

To understand the true reason of discrepancy of solutions we shall return 
to derivation of the equation (\ref{eq14}). The derivation of the equation 
(\ref{eq14}) is unusual. It is not local in the sense, that changes of the 
values relating to various points of space are considered. 
Therefore it would be possible to 
expect, that the type of symmetry of task will somehow show itself. Equation 
(\ref{eq12}) determining linear connection of infinitesimal changes of speed of 
substance and pressure in simple Riemann wave in case of spherical 
(cylindrical) symmetry is incorrect. It would be necessary to replace it 
with more general relation (see, for example, \cite{ianenko78}, 
\S~2.7 of chapter 2):
\begin{equation}
\label{eq31}
D_{ \pm}  u \pm \frac{{1}}{{\rho c}}D_{ \pm}  p \pm \frac{{\alpha cu}}{{r}} 
= 0.
\end{equation}
Where
\begin{equation}
\label{eq32}
D_{ \pm}  = \frac{{\partial} }{{\partial t}} + \left( {u \pm c} \right) 
\cdot \frac{{\partial} }{{\partial r}},
\end{equation}
 $\alpha = 0, 1, 2$ -- correspondingly, in case of plane, 
cylindrical and spherical symmetry. Thus, the equation (\ref{eq14}) for 
spherical (cylindrical) sound waves should be considered as approximated.

\section{Conclusions}
Given derivation of equation of movement of a 
solitary shock wave in the channel of variable area shows,
why this equation may be only approximate for cylindrical and spherical 
convergent waves.
But such approximation has high precision and may be used for estimations.

\end{document}